# Optical Conductivity of Twisted Bilayer Graphene Near the Magic Angle[*]

Lu Wen (文露), Zhiqiang Li (李志强), Yan He (贺言)[†]

*College of Physics, Sichuan University, Chengdu610064, China*

We theoretically study the band structure and optical conductivity of twisted bilayer graphene (TBG) near the magic angle considering the effects of lattice relaxation. We show that the optical conductivity spectrum is characterized by a series of peaks associated with the van Hove singularities in the band structure, and the peak energies evolve systematically with the twist angle. Lattice relaxation effects in TBG modify its band structure, especially the flat bands, which lead to significant shifts of the peaks in optical conductivity. These results demonstrate that spectroscopic features in the optical conductivity can serve as fingerprints for exploring the band structure, band gap and lattice relaxation in magic angle TBG as well as identifying its rotation angle.

---

[*] Project supported by the National Natural Science Foundation of China (Grant Nos.11874271 and 11874272).

[†] Corresponding author. E-mail: heyan_ctp@scu.edu.cn

## 1.Introduction

The various van der Waals heterostructures assembled from two-dimensional (2D) materials offer a unique platform combining many novel physical properties.[1-4] The high level of interest in 2D materials, combined with capabilities in device fabrication and experimental techniques, has led to an outburst of activity over the past few years. One of the simplest heterostructures is twisted bilayer graphene (TBG), which consists of two graphene layers rotationally stacked at an arbitrary angle $\theta$ to each other. In a small $\theta$, a hexagonal long-period moiré interference pattern consisting of alternating AA- and AB-stacked regions emerges, owing to a slight difference in lattice orientation. When the twist angle $\theta$ is close to the 'magic' angle near $1.10°$, [5] the electronic band structure becomes flat around the Dirac point at zero energy due to interlayer coupling.[6] The magic angle TBG has gained increased attention in theory and experiment in the last few years owing to its novel electronic properties. Recent studies on magic angle TBG have discovered correlated insulating phases with similarities to Mott insulator systems,[7] as well as strongly coupled superconducting states in ultra-flat bands close to half-filling.[8, 9] Furthermore, other intriguing quantum phenomena have been observed in magic angle TBG such as quantum anomalous Hall effect (QAHE). [10-14] These results make it possible to experimentally investigate both electronic correlations and topological phenomena in 2D van der Waals heterostructures.

Spectroscopic studies of electronic phenomena are very important to explore the physical properties in graphene-based systems.[15, 16] Optical spectrum can provide critical information about the electronic band structure,[17, 18] many-body interactions,[19, 20] quasiparticles[21] and collective excitations[22] in graphene systems. Moreover, graphene is very promising for applications in the field of optics[23] and optoelectronics[24] due to its unique mechanical, electric, magnetic and thermal properties. A multitude of exciting applications of graphene such as novel optical devices and photodetectors are being vigorously pursued by academia and industry. [25]

Previously, the optical conductivity of TBG heterostructures has also been considered. Koshino et al. [26] systematically investigated the conductivity spectra of TBG with a variety of large rotation angle $\theta$ (from $1.47°$ to $30°$). E. Bascones et al.[27] studied the effect of symmetry breaking states on the optical response, focusing on incident photon energy range below 50meV (corresponding wavelength $\lambda \approx 25$ um).

Our results highlight the optical conductivity of magic angle TBG across a wider spectra range considering the effects of lattice relaxation, which is complementary to previous reports. In addition, another notable disagreement with previous results is that our results are more directly relevant to the experiment. At present, the area with uniform twist angle $\theta$ in TBG samples is only 3-5 um in size or even smaller, due to the constraints of device fabrication and experimental technology, which is about 10 times smaller than the wavelength of incident photon (about 25 um) in Ref. [27]. Therefore, limited by the size of the TBG samples, it is difficult to compare the theoretical results of E. Bascones et al. with the experimental measurements in the short term. However, in this paper, we calculated the conductivity spectrum of TBG in the energy region from 0 to 1.0 eV and discussed in greater detail the very strong peaks in the frequency range from 0 to 0.4 eV (corresponding wavelength $\lambda \approx 3$ um) in the optical conductivity spectra. So our calculation results show the optical conductivity with higher incident energy (narrower incident wavelength), which can provide some reliable information for the experimental measurement of real TBG samples with small twist angle.

Here, we study the low-energy band structure of TBGs at a series of small rotation angles $\theta$ around the magic angle based on an effective continuum model. The optical conductivity spectra are calculated based on the Kubo formula. At such small angles, the lattice structure of TBG spontaneously relaxes to minimize the energy of the structure, leading to significant modification in band structures. The dependence of different peaks in the optical conductivity spectrum on the rotation angle $\theta$ and lattice relaxation effect is fully explored. We study in detail how the van Hove singularities in DOS are related to different interband transitions. Our results show that optical measurements are very important for studying the band structure and lattice reconstruction in TBG.

## 2. Theoretical model

A TBG superlattice is illustrated in Fig. 1. When the relative twist angle $\theta$ of the two graphene monolayers is close to the magic angle, the period of moiré superlattice is much larger than the lattice constant of graphene $a \approx 0.246$ nm, as shown in Fig. 1a, so an effective continuum model[26, 28, 29] can be used to describe the electronic band structure of TBG. Here we consider the TBG structure formed by the first and second layers of AA-stacked bilayer graphene, which rotate around the common B atom by $-\theta/2$ and $+\theta/2$, respectively. Then, the lattice constant of the moiré superlattice is $L_{\mathrm{M}} = \dfrac{a}{2\sin(\theta/2)}$. Figures 1b shows the Brillouin zone of small angle TBGs. The large hexagons in red and blue represent the first Brillouin zone of layers 1 and 2, respectively. $\boldsymbol{K}_l$ and $\boldsymbol{K}'_l$ denote the two inequivalent valleys of layer $l$ ($l = 1, 2$ corresponds to layer 1 and 2, respectively). The difference between the two $\boldsymbol{K}$ ($\boldsymbol{K}'$) wavevectors results in the mini Brillouin zone (shown as a small black hexagon), which is generated from the reciprocal lattice of the moiré superlattice. $\boldsymbol{K}_{\mathrm{s}}$, $\boldsymbol{K}'_{\mathrm{s}}$, $\boldsymbol{M}_{\mathrm{s}}$, $\boldsymbol{\Gamma}_{\mathrm{s}}$ correspond to points in the mini Brillouin zone. [30]

The effective continuum model is constructed from a continuum electronic description of massless Dirac fermions in two graphene layers that are coupled by a slowly changing periodic inter-layer hopping.[6] The states near $\boldsymbol{K}$ valley and those near $\boldsymbol{K}'$ valley are far apart in the wave space when $\theta$ is small, so the coupling between different valleys can be ignored. Near the $\boldsymbol{K}$ ($\boldsymbol{K}'$) valley, the effective low energy Hamiltonian includes the intra-layer Hamilton $H_l$ of layer $l$ and the inter-layer coupling term $U$. The continuum model can be written in a 4*4 matrix as:

$$H_{eff} = \begin{bmatrix} H_1 & U^{\dagger} \\ U & H_2 \end{bmatrix}, \tag{1}$$

with

$$H_l = -h v_{\mathrm{F}} / a (\boldsymbol{k} - \boldsymbol{K}_{\xi,l}) \cdot \boldsymbol{\sigma}, \tag{2}$$

$$U = \begin{bmatrix} u & u' \\ u' & u \end{bmatrix} + \begin{bmatrix} u & u'\omega^{-\xi} \\ u'\omega^{\xi} & u \end{bmatrix} \mathrm{e}^{\mathrm{i}\xi G_1^{\mathrm{M}} r} + \begin{bmatrix} u & u'\omega^{\xi} \\ u'\omega^{-\xi} & u \end{bmatrix} \mathrm{e}^{\mathrm{i}\xi(G_1^{\mathrm{M}}+G_2^{\mathrm{M}})r}, \tag{3}$$

in the $H_l$ term, $v_{\mathrm{F}}$ is the band velocity of the Dirac cone, $h$ is Planck constant, we take $h v_{\mathrm{F}} / a = 2.1354$ eV,[26] $\boldsymbol{\sigma}$ is the Pauli matrices, $\boldsymbol{k}$ is the two-dimensional Bloch wave vector, taken value from the moiré Brillouin zone, $\xi = \pm 1$ labels the $\boldsymbol{K}$ and $\boldsymbol{K}'$ valley, $\boldsymbol{K}_{\xi,l}$ denotes the location of the Dirac cone. In the $U$ term, $\omega = \mathrm{e}^{\mathrm{i}2\pi/3}$, $\boldsymbol{G}_l^{\mathrm{M}}$ is the reciprocal lattice vectors for the moiré superlattice, $\boldsymbol{r}$ is measured from the center of rotation in real space, $u$ and $u'$ are terms that describe the interlayer coupling.

For unrelaxed TBG systems where the two graphene layers are rigidly rotated relative to each other, the atomic registry changes continuously within the moiré period between four different stacking configurations AA, AB, BA, and saddle point(SP) as illustrated in Fig. 1a. For this case, one can use $u = u' = 110$ meV[5] in the continuum model. In practice, however, in the small-twist regime the competition between the van der Waals interlayer interaction and the elastic energy at the interface can lead to significant relaxation of the lattice structure to reduce the area of AA region with high stacking energy.[31-37] The relaxed lattice forms an array of triangular domains with alternating AB- and BA-stacking regions.[38, 39] Moreover, lattice relaxation causes corrugation in the out-of-plane direction with increased vertical separation between the AA regions compared to that in AB regions.[31] The effects of lattice relaxation can be taken into account by using $u = 0.0797$ eV and $u' = 0.0975$ eV[29] in the effective continuum model as shown previously.

Using the eigen–wave functions obtained from the effective continuum model, Eq. (1), the optical conductivity $\sigma(\omega)$ of TBG can be calculated by the Kubo formula:

$$\sigma(\omega) = \frac{e^2\hbar}{iS} \sum_{m,n} \frac{f(E_m) - f(E_n)}{E_m - E_n} \frac{\left|\langle a_m | v_x | a_n \rangle\right|^2}{\hbar\omega + i\eta - E_n + E_m}, \qquad (4)$$

here $e$ represents the elementary charge, $\hbar$ is reduced Planck constant, $S$ is the area of the system in real space, $f(E)$ is the Fermi distribution function, $E_m(E_n)$ and $|a_m\rangle(|a_n\rangle)$ are the eigenenergy and eigenstate of the system obtained by solving the Hamiltonian, $v_x$ represents the velocity operator, $\eta$ is the phenomenological broadening, which is 0.003 eV,[26] $\hbar\omega$ is the incident photon energy.

## 3. Results and discussion

Based on the effective continuum model described above, we calculate the electronic band structure of relaxed (black solid lines) and unrelaxed (red dashed lines) TBGs at three rotation angles $\theta$ (1.05°, 1.10° and 1.16°, respectively) close to the magic angle, as shown in Figure 2. Here, the path of band structures is along the line $\boldsymbol{K}_s - \boldsymbol{\Gamma}_s - \boldsymbol{M}_s - \boldsymbol{K}'_s$ in the moiré Brillouin zone scheme, and the origin of the band energy axis (the band energy $E = 0$ eV) is set to the charge neutral point. Regardless of lattice reconstruction, a set of flat bands appears at the Dirac point due to the competition between the kinetic energy and the interlayer hybridization energy. For unreconstructed TBG, the flat bands are confined to less than about 15 meV, and the bandwidth decreases with the increasement of twist angle $\theta$. However, under the influence of lattice relaxation, the flat bands near the Dirac point are obviously gapped from the high energy bands in both the electron side and the hole side. The size of such gaps is dependent on the twist angle $\theta$, which appears to be minimum (about 10 meV) at $\theta = 1.05°$ among the three rotation angles (Fig. 2a). In addition, the lattice reconstruction also has a great influence on band velocity in the flat bands. The lowest energy bands at the Dirac point in reconstructed TBGs are significantly flattened compared to the unrelaxed TBGs at the same angle.

In Fig. 3, we plot the optical conductivity $\sigma(\omega)$ of TBGs at various rotation angles in a wide frequency range, when the temperature is close to 0 K. The red dashed line and black solid line represent the conductivity of unreconstructed and reconstruced TBGs, respectively. The conductivity is plotted in units of $\sigma_{mono} = \dfrac{g_v g_s}{16} \dfrac{e^2}{\hbar}$, which is the dynamical conductivity of monolayer graphene, and the spin and valley degeneracy are $g_s = g_v = 2$, respectively.[40, 41] As shown in Fig. 3a-c, the conductivity spectrum exhibits a series of characteristic peaks, mainly including the first group of peaks around 0.1 eV and the second group of peaks in the frequency range from 0.2 eV to 0.4 eV. The peak structure almost systematically evolves with the rotation angle $\theta$. When the effect of lattice relaxation is taken into account, the first group of peaks move to higher energies, and the peaks in the second group also shift significantly compared to the unreconstructed TBG system. The first peak in $\sigma(\omega)$ in reconstruced TBGs corresponds to the interband transition from the flat band to the higher band, which can provide an effective experimental way to determine the gap between the flat bands and other bands under lattice relaxation. These gaps are very important for understand novel transport properties of magic angle TBG. Therefore, experimental measurements of the optical conductivity of TBG can provide insights into the lattice relaxation effects in this system.

Fig. 4 shows the band structure, DOS and the optical conductivity of relaxed TBG with $\theta = 1.05°$. The black solid line (red dashed line) represents the energy bands of $K$ ($K'$) valleys in Fig. 4a. In Fig. 4a and 4b, each characteristic saddle point of the band structure is corresponding to a van Hove singularity in the DOS. The sharp peaks in the optical conductivity spectrum (Fig. 4c) are associated with optical transitions involving the van Hove singularities in the DOS (Fig. 4b). For instance, a peak numbered ① near $E = 0.02$ eV in Fig. 4c corresponds to the interband transition from the highest valence band to the second conduction band, as indicated by the magenta arrow in Fig. 4a. Moreover, the major peaks labeled ①-⑦ in the optical conductivity in Fig. 4c can be assigned to transitions between different bands labeled by arrows in Fig. 4a. There are also unmarked interband transitions from the other valence bands to the flat band in Fig. 4a, which are similar to transitions indicated by red arrows due to the electron-hole symmetry in the effecitive continuum model.

Fig. 5a shows the optical conductivity of TBG with rotation angle $\theta = 1.05°$, in the presence of lattice relaxation, for different values of the electronic filling. The lower panel (Fig. 5b) is the enlarged plot of the energy region from 0 to 0.1 eV in Fig. 5a. The black solid line and red dashed line represent the conductivity for the Fermi energy level at the charge neutrality point (CNP) and full (or empty) flat bands, respectively. As shown in Fig. 5a, at high energies, the peak energies are similar for different values of the electronic filling in the conductivity spectra. Remarkably, in Fig. 5b, the first group of peaks for full (or empty) flat bands move significantly towards lower energies, and at the same time, the Drude peak at zero frequency disappears compared to the chemical potential of the undoped system $\mu_{CNP}$. Next, to fully understand the dependence of optical conductivity on rotation angle $\theta$, we additionally calculated the conductivity of relaxed TBG with rotation angle $\theta = 1.20°$ and $0.80°$ in Fig. 5c, and larger angles were described in detail in Ref. [26]. Fig. 5c shows that the peak energies evolve systematically with the twist angle θ, so they can be used as fingerprints to identify the twist angle. Especially, for the conductivity of TBG with θ = 0.80°, the first group of peaks degenerate into only one strong peak relative to that of TBGs with other larger twist angles.

## 4. Conclusion

We have theoretically calculated the electronic band structure and optical conductivity spectrum of TBGs at various rotation angles in the vicinity of magic angle, and explored the effects of lattice relaxation. We have shown that a series of characteristic peaks in the conductivity spectrum is related to the van Hove singularities in the DOS, which enables us to extract a lot of important information about interband transitions and band structure. When lattice relaxation effects exist, the gaps between the flat bands and other bands in magic angle TBG can be quantitatively studied according to the lowest energy peak in the conductivity spectrum by optical measurements. Such gaps are very important for understand the novel transport properties of magic angle TBG. In addition, the peaks in conductivity spectrum shift significantly under the effect of lattice reconstruction. Our study paves the way for exploring the physics of TBG employing optical techniques.

## Figures

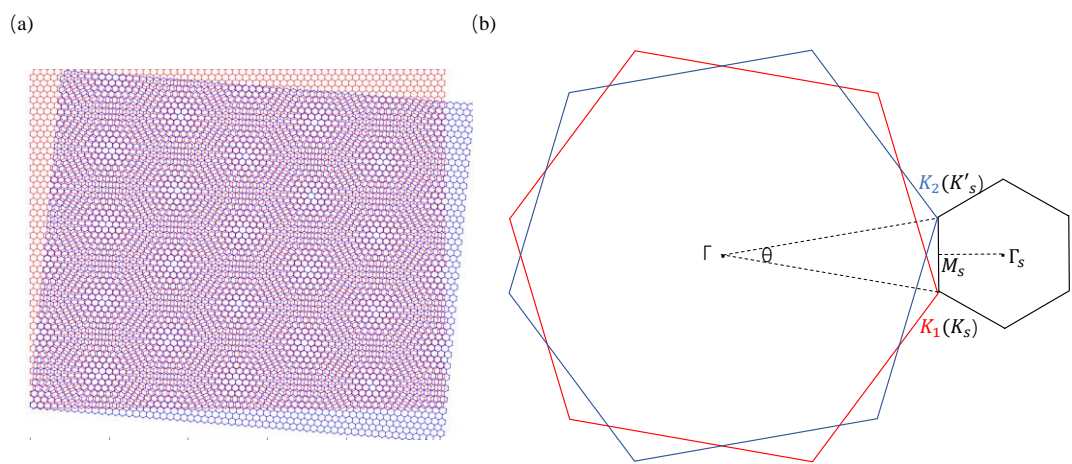

Fig.1. (Color online) (a) The moiré pattern of twisted bilayer graphene (TBG). The moiré wavelength is $L_M = \dfrac{a}{2\sin(\theta/2)}$, where $a = 0.246$ nm is the lattice constant of graphene and $\theta$ is the rotation angle. (b) The moiré Brillouin zone (Small black hexagon) is constructed from the difference between the two $K$ ($K'$) wavevectors for the two layers (Red hexagon:layer 1, blue hexagon: layer 2). $K_s$, $K'_s$, $M_s$, $\Gamma_s$ correspond to points in the moiré Brillouin zone.

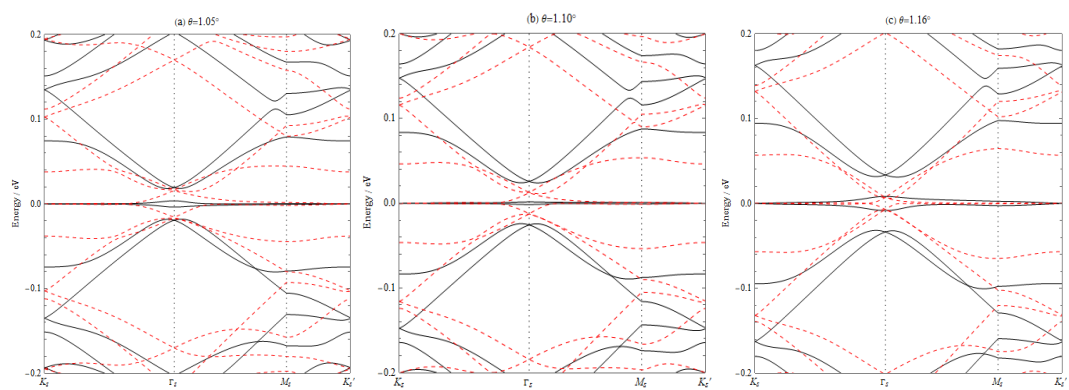

Fig. 2. (Color online) Band structure of relaxed (black solid lines) and unrelaxed (red dashed lines) TBGs at various rotation angles closed to the magic angle near 1.10 ° in (a) 1.05 °, (b) 1.10 ° and (c) 1.16 °, respectively. The path of band structures is along the line $\boldsymbol{K}_s - \boldsymbol{\Gamma}_s - \boldsymbol{M}_s - \boldsymbol{K'}_s$ in the moiré Brillouin zone scheme, and the Dirac point energy is set to 0 eV.

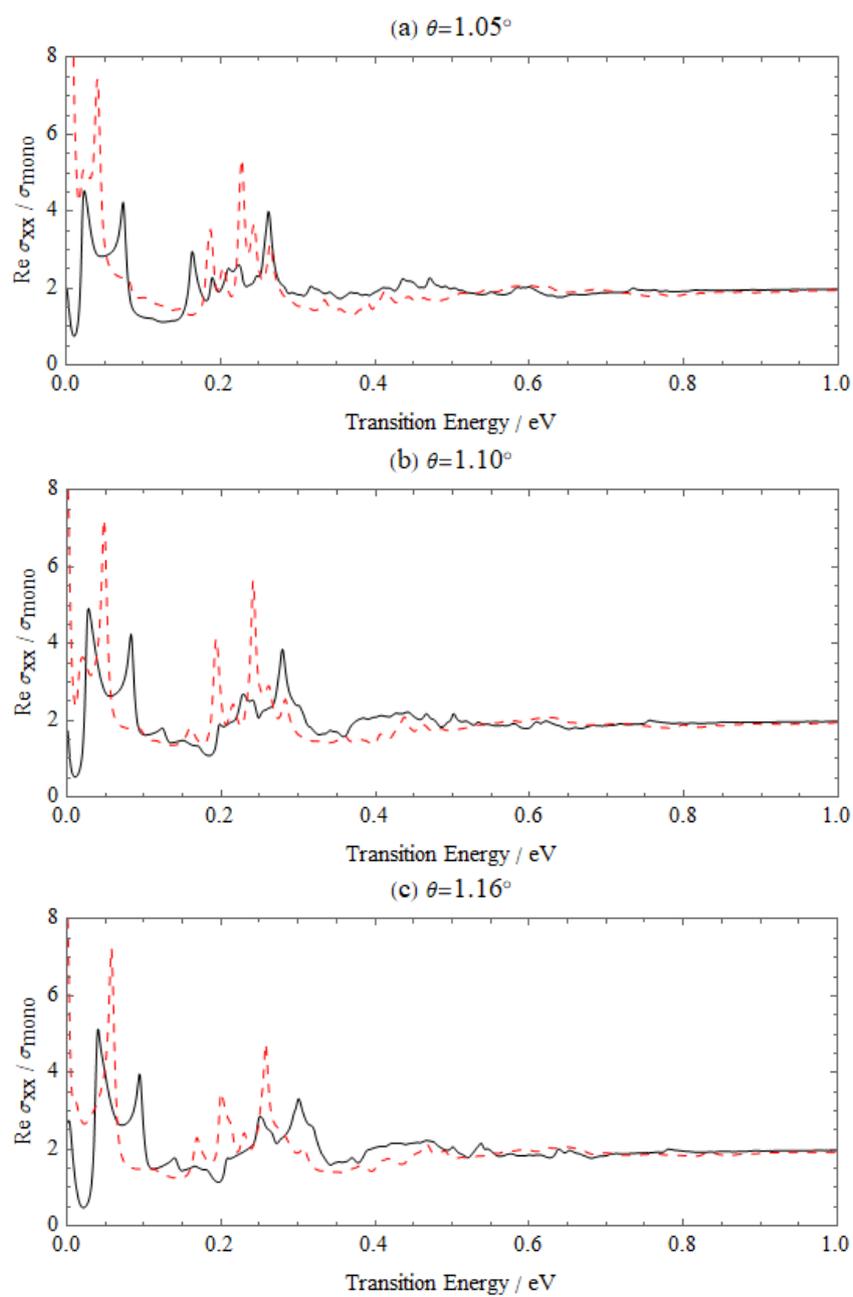

Fig. 3. (Color online) The conductivity spectrum of relaxed (black solid lines) and unrelaxed (red dashed lines) TBGs with (a) $\theta = 1.05\,°$, (b) $\theta = 1.10\,°$ and (c) $\theta = 1.16\,°$.

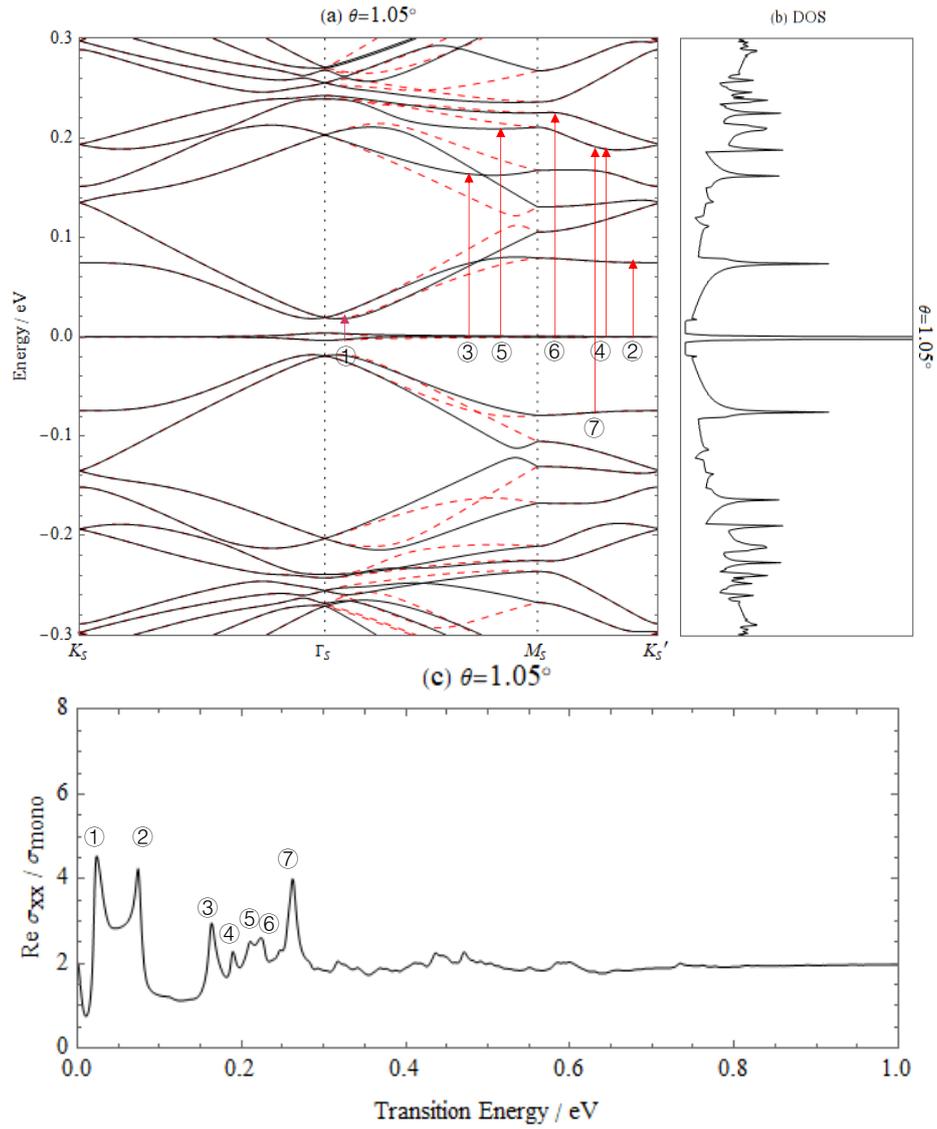

Fig. 4. (Color online) (a) Band structure, (b) density of states (DOS) and (c) conductivity spectrum of relaxed TBG with $\theta = 1.05\,°$. The black solid line (red dashed line) represents the energy bands of $K$ ($K'$) valleys. Red solid arrows labeled ①-⑦ in (a) represent interband transitions corresponding to the major peaks numbered ①-⑦ in the conductivity spectrum in (c).

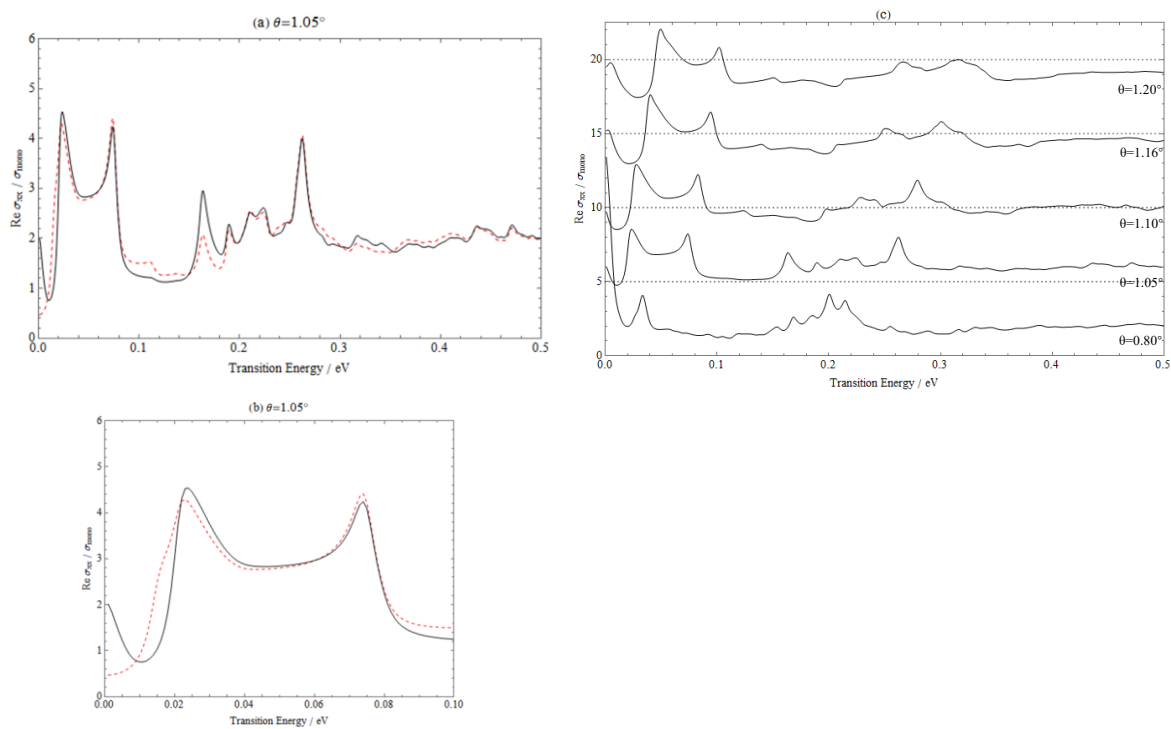

Fig. 5. (Color online) (a) The optical conductivity of relaxed TBG at $\theta = 1.05\,°$ for the Fermi energy level at the CNP (black solid lines) and full (or empty) flat bands (red dashed lines). (b) The enlarged plot of the energy region from 0 to 0.1 eV in Fig. 5a. (c) The optical conductivity of relaxed TBGs with various rotation angles. (The curves are relatively shifted to make the plot clear.)